\documentclass[11pt,twoside]{article}

\usepackage{asp2006}
\usepackage{epsf}
\usepackage{psfig}
\usepackage{lscape}
\usepackage{wrapfig}

\begin{document}
\title{The Role of MHD in the ICM and its Interactions with AGN Outflows}   %%% Fill in title
\author{T. W. Jones}   %%% Fill in author names
\affil{University of Minnesota, Department of Astronomy, Minneapolis, MN 55455}    %%% Fill in author affiliations

\begin{abstract} %%% Abstract to run on from here.
Magnetic fields probably play a central role in the dynamics and
thermodynamics of ICMs and their interactions with AGNs,
despite the fact that the fields usually contribute
relatively little pressure; i.e., the ICM is a ``high-$\beta$'' plasma. 
More typically, the roles of magnetic fields
come through ``microscopic'' influences on charged
particle behaviors, and through magnetic tension,
which can still be significant in subsonic, high-$\beta$ flows.
I briefly review these issues, while exploring the underlying
question of using the commonly-applied magnetohydrodynamics
model in the ICM when Coulomb scattering mean free paths can
sometimes exceed tens of kiloparsecs.
\end{abstract}

%%% MAIN BODY OF TEXT GOES HERE. CONSULT "INSTRUCTIONS FOR AUTHORS USING
%%% LATEX2E MARKUP", SECTIONS 2.3-2.6 FOR HELP WITH EQUATIONS, FIGURES,
%%% AND TABLES.

%\section{}   %%% Top level section head (remove "%" symbol)
\section{Introduction}   
Powerful AGN jets typically deposit most of their momentum and energy in
low density, fully ionized and largely collisionless plasmas
that constitute the intracluster media, or ICMs.
Inevitably in such media, charge mobility differences lead to electrical 
currents and, thus, to magnetic fields. 
Those magnetic fields limit the motions of the charged particles,
affecting the momentum, energy and charge transport through the
plasmas and the electric currents establishing
the magnetic fields themselves. The 
detailed physics of these interactions is complex and most accurately
explored on ``microscopic'' levels through the tools of plasma physics. 
On the other hand the vast degrees of freedom inherent in plasma
treatments, especially when applied on ``macroscopic'' scales, often 
makes such treatments unwieldy. It is common, instead to model 
these media, including their interactions with AGN jets,
through magnetohydrodynamics (MHD). Like other
continuum approximations, MHD carries with it assumptions
that can potentially mask or exclude relevant physics. On the other hand,
when appropriate, MHD provides a very powerful
tool, and sometimes the only practical tool for the exploration of
the interactions at the center of discussion in
this meeting. It turns out that even nominally
weak magnetic fields can have profound influences on
both flow dynamics and thermodynamics in these settings.
I was asked to explore these issues in this presentation.

I begin with a short review of the assumptions built into 
MHD and an evaluation of their applicability in ICMs.
Concluding, with some caveats, that the MHD model is appropriate on
many important length and time scales in clusters I summarize some
common dynamical and thermodynamical MHD issues
and then discuss
some specific AGN/cluster interactions where the presence of
magnetic fields appear to be important. 
My brief comments are necessarily incomplete. 
More thorough discussions of MHD and its 
connections to plasma physics are widely available
\citep[e.g.,][]{boyd03, goed04, kuls04, priest00}.

I do not consider the origins of magnetic fields in clusters or
radio galaxies, although that is of much current interest.
The proceedings of the conference, 
``The Origin and Evolution
of Cosmic Magnetic Fields'', provide a good introduction to many
aspects of the problem \cite{beck}. Suffice it to say 
that both observational and theoretical estimates of
magnetic field strengths in clusters are typically
in the vicinity of a few microGauss, give or take an order of magnitude
\cite[e.g.,][]{car02, dolag06, kron96, ryu07, schek05, vik01};
similar, if somewhat larger, estimates apply in the lobes of FRI and 
FRII radio galaxies \cite[e.g.,][]{croston05, worr00}. 
I will take a microGauss for a characteristic magnetic field 
strength. Additionally, 1 keV and
$10^{-2}~{\rm cm}^{-3}$ provide convenient
fiducial cluster temperature and particle density values. 
In addition I assume the ICM is pure
hydrogen and that protons and electrons share the same temperature.

\section{How well does MHD apply to the ICM?}

MHD is widely applied to
the dynamics of the ICM. 
Being a single-fluid, continuum model, MHD requires
on timescales of dynamical
interest, $t_d \sim \ell_d/u_d$, that the particle populations 
pass through a sequence of local equilibria; i.e.,  effective particle
interaction times, $t_c\ll t_d$,
and interaction lengths, $\ell_c \ll \ell_d$, where $\ell_d$ and $u_d$ are
representative dynamical lengths and speeds.
In addition MHD requires relevant charged particles to be effectively magnetized; 
that is, their gyroradii, $r_g = v_{\perp}mc/(eB)\ll \ell_d$ and gyroperiods, $t_g = 2\pi r_g/v_{\perp} = 2\pi/\omega_c \ll t_d$.
MHD usually neglects the displacement
current in Ampere's Law, a valid approximation if relevant wave speeds
are nonrelativistic; i.e., $\omega \ll k c$. 
Standard MHD also
assumes isotropic (scalar) pressure and transport 
properties for the fluid, global charge neutrality. It neglects 
electrostatic forces due to local charge fluctuations, allowing the
fields and bulk fluid to vary simultaneously.
There are variants to the standard MHD model that
relax various of these assumptions, such as a single fluid, pressure 
and
transport isotropy \cite[e.g.,][]{brag65, boyd03, goed04, kuls04},
although they are more complicated to apply.

The appropriate interaction lengths are central to 
this discussion. Beyond their role in validating the continuum dynamics
model itself they also control basic fluid
properties needed in the MHD model, including 
transport behaviors such as viscosity, $\nu$,
thermal conductivity, $\kappa$, and electrical resistivity, $\eta$.
The standard kinetic theory expressions for these transport
coefficients are \cite[e.g.,][]{boyd03} $\nu = k_B T t_{c,p}/m_p$,
$\kappa = (5 k_B)/(2m_e) n k_B T t_{c,e}$ and $\eta = m_e c^2/(n e^2 t_{c,e})$,
where the second subscript, `p' or `e' refers to protons or electrons.
When $t_c$ is the Coulomb collision time discussed below
the thermal conductivity is known as the ``Spitzer conductivity'',
while the analogous viscosity is called the ``Braginskii viscosity''.

Unlike laboratory fluids where strong binary collisions
typically establish equilibria,
the hot, rarefied and fully ionized conditions in an ICM lead to a dominance by
the combined effects of many weak 
interactions. The number of charged 
particles participating in Coulomb
collisions with a proton or electron is determined by the number of 
particles, $N_D$, found inside the so-called Debye sphere of radius, $\lambda_D$, given by
\begin{equation}
\lambda_D = [k_B T /(4\pi n e^2)]^{1/2} = 
[v_{th}^2/(3\omega_p^2)]^{1/2} = 
2.4\times 10^5 (T_{keV}/n_{-2})^{1/2}~{\rm cm},
\label{lambda}
\end{equation}
where 
\begin{equation}
v_{th} = \sqrt{3k_B T/m} =  5.4\times 10^7 T_{keV}^{1/2}~(m_p/m)^{1/2} ~{\rm cm/s}
\label{vth}
\end{equation}
is the characteristic particle thermal speed
and 
\begin{equation}
\omega_p = \sqrt{4\pi n e^2/m}= 1.3\times 10^2~(m_p/m)^{1/2}  n_{-2}^{1/2}~{\rm s}^{-1}
\label{omegap}
\end{equation}
is the plasma frequency for
that species. In the numerical expressions, $m_p$ is the proton mass
and the plasma density is expressed in units $n = 10^{-2} n_{-2}~{\rm cm}^{-3}$.
For electrons, 
$v_{th} = 2.3\times 10^9 T_{keV}^{1/2}~{\rm cm/s}$ and 
$\omega_p = 5.6\times 10^3 n_{-2}^{1/2}~{\rm s}^{-1}$.
When $N_D \gg 1$ the Coulomb scattering cross section is enhanced
over that for strong binary collisions,
$\sigma_c \sim 4\pi e^4/(3k_B T)^2$, 
by a factor proportional to $\ln{(3N_D)} \equiv \ln{\Lambda}$ due to random, 
thermal fluctuations in charge density.

In clusters 
$N_D = (4\pi/3)n_e \lambda_D^3 \approx 5.4\times 10^{14} 
(T_{keV}/n_{-2})^{1/2}\gg 1$, and $\ln{\Lambda} \sim 30 - 40$.
Using $\ln{\Lambda} = 35$, the effective proton collision time is then \cite{spitz62} 
\begin{equation}
t_{c,p} \approx 4.1\times10^4 T_{keV}^{3/2}/n_{-2}~{\rm yrs},
\label{tc}
\end{equation}
while the electron collision time, $t_{c,e}$, is smaller by a factor
$(m_e/m_p)^{1/2}$. This gives both particle species a
collisional mean free path , $\ell_c = v_{th} t_c$, given by
\begin{equation}
\ell_{c,p,e} \sim 22~ T_{keV}^{5/2}/n_{-2}~{\rm pc}. 
\label{lc}
\end{equation}

Temperature equilibrium between protons and electrons actually requires
more stringent constraints, since the energy exchange during e-p collisions
is proportional to $m_e/m_p$. Consequently the time for thermal equilibration 
between the two species is $\sim (m_p/m_e)^{1/2} t_{c,p}$.
In relatively denser and cooler ICMs these various Coulomb
times
and lengths should be short enough to support a fluid model on many scales of
interest. However, in less dense environments
outside cores and particularly in hotter ICMs Coulomb interactions 
can be uncomfortably slow in
this context. With $n = 10^{-3}~{\rm cm}^{-3}$ and $T = 10 keV$ 
we have from
equations \ref{tc} and \ref{lc} $t_{c,p} \sim 10^7~{\rm yrs}$ and $\ell_c \sim 10^2~{\rm kpc}$,
for instance. 

On the other hand, except for uniform, static and unmagnetized  media,
Coulomb collisions probably underestimate the 
effective interactions available. As a starter, velocity space
plasma instabilities in a dynamical setting  may redistribute particle motions 
more rapidly than Coulomb collisions.
For example, if there are density or velocity gradients on scales smaller than
the Coulomb interaction lengths, particles should
stream against these gradients. Then
the so-called two-stream instability can
come into play. The two-stream instability
leads to particle bunching and associated coherent electrostatic
fields on scales of the Debye length that can redistribute particle
momenta. In the simplest case two equal, cold, like-charged 
beams interpenetrate with a relative speed $v_b$. 
Fluctuations with wavenumber $k \sim \omega_p/v_b$
are unstable with very fast growth rates 
$\Gamma ~\sim \omega_p$ \cite[e.g.,][]{stix92}.
If we associate $v_b$ with the eventual thermal speed, then $k\lambda_D \sim 1$,
and we see that particle motions are redistributed very quickly
on scales of the Debye length. More realistically only a fraction of
the total particle population would be involved in streaming,
so growth rates would be reduced proportionately.
Still, the coherent nature of the induced charge
fluctuations can substantially boost the
effective scattering rate compared to incoherent Coulomb scattering. 

Magnetic fields obviously influence this picture substantially, as well,
since proton and, especially, electron gyroradii, $r_g$, and 
gyroperiods, $t_g = 2\pi/\omega_g$, in the ICM
should be small; in particular, $r_g/ \ell_c \ll 1$
or equivalently, 
$t_c \omega_g \gg 1$. 
For thermal protons and electrons, respectively, 
$r_{g,p} = 4.6\times 10^9 T_{keV}^{1/2}/B_{\mu G}~{\rm cm}$,
$~r_{g,e} = 1.0\times 10^8 T_{keV}^{1/2}/B_{\mu G}~{\rm cm}$,
and with $\omega_g = eB/(mc)$,
$~\omega_{g,p} = 0.01 B_{\mu G}~{\rm sec}^{-1}$,$~\omega_{g,e} = 18 B_{\mu G}~{\rm sec}^{-1}$.
Accordingly, 
\begin{equation}
t_{c,p}\omega_{g,p} = (m_e/m_p)^{1/2} t_{c,e} \omega_{g,e} 
\approx 1.2\times 10^{10} T_{keV}^{3/2} B_{\mu G} n_{-2}^{-1},
\label{tcp}
\end{equation}
which will generally be large in the ICM. This will guarantee, as well,
that the ICM is magnetized sufficiently to apply MHD, whenever $\ell_c \ll \ell_d$.

MHD assumes charge quasi-neutrality, so that electrostatic
fields can be neglected. In effect, when there is global
neutrality one requires inside fluctuations that $|q|/(n e)\ll 1$,
where $q$ is the local charge density. 
Dimensional analysis combining the equation for charge
continuity, $\partial q/\partial t = - \nabla\cdot j$, with Ampere's
law and the previously mentioned nonrelativistic phase speed 
constraint leads to the condition 
$\omega/\omega_{g,e} \ll (\omega_{p,e}/\omega_{g,e})^2 \approx 10~n_{-2}/B_{\mu G}^2$,
where 
the condition is applied to electrons
because of their greater mobility. This limits us to ICM fluctuations
roughly slower than a gyroperiod, so that free electrons can
redistribute themselves to short out local electrostatic fields.

In a quasiuniform field, particles spiral along the field
with orbital radius, $r_g$, traversing a longitudinal distance, $\ell_c \sim v_{||} t_c$
before scattering substantially changes their pitch angles.
Scattering also introduces a transverse gyrocenter shift $\sim r_g$,
so in the limit $r_g/\ell_c \ll 1$, the relative
transverse and parallel particle diffusion rates would be
$\sim (r_g/\ell_c)^2$. 

Thermal energy and also momentum transport across fields are controlled by
cross-field diffusion, which reduces by similar factors, $\sim (r_g/\ell_c)^2$,
the thermal conductivity transverse to a uniform field and also the viscosity in 
response to velocity gradients transverse to B  \cite{brag65, spitz62}.
The coefficients parallel to $B$ are also modified by dimensionality
influences, as is the electrical
resistivity tensor, by factors of order unity \cite[e.g.,][]{kuls04}.

However, the ICM is thought to be turbulent with turbulent
motions contributing perhaps $\sim 10 \%$
of the total pressure \cite[e.g.,][]{schueck04}. In that case
the magnetic field might not be at all uniform. Spatial diffusion then could be
limited by the scale for bending of field lines or wandering of 
individual field lines. There has been much discussion of
this topic recently, especially with regard to thermal
conductivity \cite[e.g.,][]{nar01} and viscosity \cite[e.g.,][]{rey05}. 
Lazarian has recently provided a nice
outline of the issues \cite{laz07}. 

For our discussion the relevant points would
be these. The character of the turbulence depends on whether 
$u_d(\ell)/v_A$ is greater or lesser than unity, 
where
\begin{equation}
v_A = B/(4\pi n m_p)^{1/2} \approx 22 B_{\mu G} n_{-2}^{-1/2}~{\rm km/s}
\label{alfven}
\end{equation}
is the 
Alfv\'en velocity and $u_d(\ell)$ is the turbulent velocity on scales, $\ell$.
When  $u_d(\ell)/v_A$ is large the turbulence is isotropic and Kolmogorov,
so that $u_d(\ell) \propto \ell^{1/3}$. When $u_d(\ell)/v_A$ is small,
on the other hand, field line tension is sufficient to resist
bending, so that turbulent motions and magnetic field structures become
decidedly anisotropic. On the turbulence injection scale, $L$, in 
clusters the condition $u_d(L)/v_A\gg 1$
probably applies under most circumstances. 
The transition to anisotropic, MHD
turbulence then occurs below a scale 
\begin{equation}
\ell_A/L \sim 10^{-2}B_{\mu G}^3/(u_{L,100}^3 n_{-2}^{3/2}), 
\end{equation}
where $u_{L,100}$ is the injection-scale turbulent velocity 
in units of 100 km/s. Particles should free stream
no farther than the lesser of $\ell_c$ and $\ell_A$. On
scales larger than $\ell_A$ the particle diffusion (and
the associated transfer coefficients) should be isotropic
with 
\begin{equation}
\kappa \sim (1/3) min(1,\ell_A/\ell_c) \kappa_{Spitzer} \le (1/3) \kappa_{Spitzer}
\end{equation}
and 
\begin{equation}
\nu \sim (1/5)  min(1,\ell_A/\ell_c) \nu_{Braginskii} \le (1/5) \nu_{Braginskii}
\end{equation}
\cite[e.g.,][]{gruz02, gruz06}, 
assuming Coulomb scattering. From the relations already given
we can estimate that 
\begin{equation}
\ell_A/\ell_c \sim 0.4 B_{\mu G}^3 T_{keV}^{-5/2} n_{-2}^{-3/2} u_{L,100}^{-3} L_{kpc}.
\end{equation}
For common ICM conditions $\ell_A/\ell_c < 1$
and sometimes $\ell_A/\ell_c \ll 1$, giving viscosities
and the thermal conductivities at least several times smaller and
perhaps much smaller than Coulomb values.

Finally, standard MHD assumes an isotropic plasma pressure.
To understand this constraint consider particles in a smoothly varying 
magnetic field. Without collisions the particles move adiabatically,
conserving $v_{\perp}^2/B$ and, under incompressible conditions, $v_{||}B$. Again
assuming incompressible changes and defining a pressure anisotropy, 
$A = (P_{\perp} - P_{||})/(2 P_{||})$, with $P = (1/3)n v^2$, it is easy to
show starting from $A = 0$ that $dA/dt \approx (3/2) d ln(B)/dt $ \cite{hall80}. 
This growth in pressure anisotropy is limited to the collision time.
So, if we choose $t_d$ for changes in the magnetic field,
we have $A \sim (3/2) t_c/t_d$. Thus, where 
collisions are fast on dynamical times
the pressure anisotropies, $A$, should be small, and a scalar
pressure representation adequate. Firehose and mirror
instabilities that develop easily on scales down to a few
gyroradii from pressure anisotropies when
magnetic fields are weak \cite[e.g.,][]{hall80} also will tend to reduce 
anisotropies, while enhancing magnetic field irregularities.

We can summarize this section by saying that if we must depend entirely
on Coulomb scattering, then interaction lengths may be sufficient in
cooler and denser ICMs to support the MHD model, but not so in hotter and more
rarefied ICMs. On the other hand plasma processes and even
weak magnetic fields may substantially change that situation, allowing
MHD to be a meaningful model at least beyond kiloparsec scales
in most ICM environments of interest.

\section{Ideal MHD?}

Accepting some caveats from the previous section, we can proceed and
express the standard dissipative MHD equations in terms of
an isotropic gas pressure, $P$, and scalar viscosity, $\nu$,
thermal conductivity, $\kappa$, and resistivity, $\eta$
in electromagnetic units as
\begin{equation}
\frac{\partial\rho}{\partial t} + {\nabla}\cdot
\left(\rho { u}\right) = 0,
\label{cont}
\end{equation}
\begin{equation}
\frac{\partial u}{\partial t} + { u}\cdot{\nabla}
{ u} +\frac{1}{\rho}{\nabla}P - \frac{1}{4\pi\rho}\left(
{\nabla}\times{ B}\right)\times{ B} = \nu\nabla^2 u,
\label{navier}
\end{equation}
\begin{equation}
{\partial P\over\partial t} + { u}\cdot{\nabla}P
+ \gamma P{\nabla}\cdot{ u} = \left(\gamma-1\right)
[\nabla\cdot(\kappa\nabla T) + \frac{\eta}{(4\pi)^2}(\nabla\times B)^2],
\label{energy}
\end{equation}
\begin{equation}
{\partial{ B}\over\partial t} - {\nabla}\times\left(
{ u}\times{ B}\right) = \frac{\eta}{4\pi}\nabla^2{ B},
\label{induct}
\end{equation}
\begin{equation}
\nabla\cdot B = 0.
\label{divb}
\end{equation}

The equations represent conservation of mass (eq. \ref{cont}), momentum
(eq. \ref{navier}), energy (eq. \ref{energy}) and magnetic flux (eq. \ref{divb})
plus Faraday's induction law  combined with Ohm's law (eq. \ref{induct}).
It is most common to apply the ideal, or 
nondissipative version of these equations,
in which the viscosity, resistivity and thermal conductivity are neglected.
How appropriate is ideal MHD in the ICM?

Electrical resistivity is the easiest to evaluate. A comparison
of the inductive and resistive terms in eq. \ref{induct} leads to the
condition that resistive dissipation and field diffusion can
be neglected when the magnetic Reynolds number, 
$R_M = u_d \ell_d/\eta \gg 1$. 
Assuming Coulomb collisions,
\begin{equation}
R_M \sim 2.3\times 10^{20} u_{d,100}\ell_{d,kpc} T_{keV}^{3/2},
\end{equation}
so that, as expected, the
magnetic field is nicely frozen-in on most scales of interest. Ohmic heating
can be neglected in eq. \ref{energy}
when $R_M \beta \gg 1$, where $\beta = P/P_B$, with $P_B = B^2/(8\pi)$.
We have $\beta \approx 400 n_{-2}T_{keV}/B_{\mu G}^2 \gg 1$,
so that Ohmic heating is generally unimportant.

Thermal conduction in eq. \ref{energy} can be neglected with respect
to convective heat transport when
the Peclet number, 
$P_e = u_d \ell_d P/(\kappa T) \gg 1$. With Spitzer conductivity we
have 
$\kappa \approx 2.6\times 10^{12} T_{keV}^{5/2}~{\rm erg/(cm~s~K)}$, 
so that
\begin{equation}
P_e \sim 3.3\times 10^{-2} u_{d,100}\ell_{d,kpc}n_{-2} T_{keV}^{-5/2}.
\end{equation}
Fig. 1 illustrates $P_e$ from this expression.
It is clear,
as many have noted before \cite[e.g.,][]{bert86, nar01, fab05, pope05} that
Spitzer conductivity would not be ignorable in the ICM. However, as 
discussed above, especially if the
magnetic field is tangled on scales small compared to the
Coulomb scattering length
the effective conduction could be substantially smaller than Spitzer.
How small remains an open question.
Cold fronts, which are contact discontinuities sometimes seen
to exhibit sharp temperature changes,
even on scales less than a Coulomb scattering length \cite{vik01},
demand a much smaller conductivity, at least locally.

\begin{wrapfigure}[22]{r}{6.3 cm}
\vspace{5.7 cm}
\includegraphics{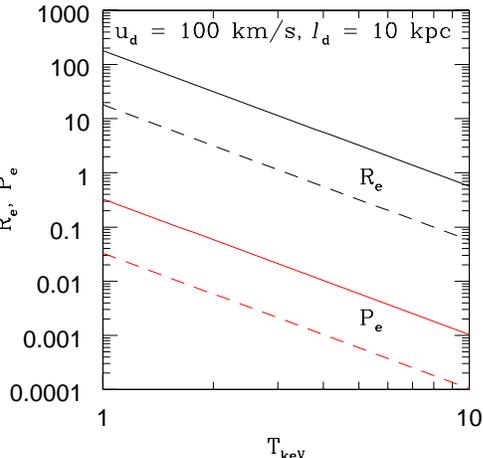}
\caption{Reynolds number, $R_e$, (upper, black lines) and 
Peclet number, $P_e$, (lower, red lines) assuming
Braginskii viscosity and Spitzer conductivity. Densities are
$n = 10^{-2}~{\rm cm}^{-3}$ (solid) and $n = 10^{-3}~{\rm cm}^{-3}$ (dashed).}
\end{wrapfigure}
The importance of viscosity is measured by the relative size of
the inertial to the viscous terms in eq. \ref{navier}; that is, by the Reynolds number,
$R_e = u_d \ell_d/\nu$. When $R_e \gg 1$ inviscid dynamics is a good approximation.
The Reynolds number must also be large for turbulence to be established
for scales, $\ell_d$ and $u_d$.
As for thermal conduction, it has been pointed out that Coulomb (Braginskii) 
viscosity
can sometimes be too large in the ICM to assume these limits \cite[e.g.,][]{rey05}.
Here 
\begin{equation}
R_e \approx 18 u_{d,100} \ell_{d,kpc} n_{-2} T_{keV}^{-5/2}.
\end{equation}
$R_e$ and $P_e$ are related by the Prandtl number,
$P_r = P_e/R_e \approx 0.08 (m_e/m_p)^{1/2}$, reflecting the ion diffusion
and electron diffusion origins for viscosity and thermal conduction.

Fig 1. illustrates the Braginskii expression for $R_e$.
As before, in cooler, denser ICM environments this estimate often gives
$R_e \gg 1$, whereas
in hotter rarefied ICMs that is not the case. One would have to reduce the viscosity
two orders of magnitude below the Braginskii value to establish
$\le 100$ km/s turbulence on 10 kpc scales if $T_{keV} = 5$ and $n_{-2} = 0.1$, 
for instance. However, if turbulence is somehow established down to scales
below $\ell_c(Coulomb)$, perhaps initially aided by plasma processes, 
the effective particle streaming lengths may become
small compared to $\ell_c$, allowing the  large $R_e$ needed for
turbulent conditions to be maintained.
Given observations that suggest
turbulence in ICMs on scales of a few kpc and
recognizing that radio halos may depend on turbulent cosmic ray
reacceleration from small eddies \cite{brun07}
it is perhaps reasonable to 
apply inviscid MHD on those scales. We may hope that ultimate 
clarification of this issue will come from definitive measures of
turbulence in a wide variety of ICM settings, especially on the 
smallest detectable scales.

\section{MHD influences on AGN/ICM interactions}

I assume thus, with some
reservations, quasi-ideal versions of eqs. \ref{cont}-\ref{induct}.
Our primary interest, in any case, is the
dynamical influence of the field, as expressed by the Lorentz
force term in eq. \ref{navier}. The standard criterion for evaluating
its importance
is to compare the magnetic pressure portion of the Lorentz force
to the gas pressure gradient, so that
magnetic effects are assumed small
whenever $\beta >> 1$. In the ICM we have
$\beta \approx 400 n_{-2}T_{keV}/B_{\mu G}^2$.
Using this criterion alone we would conclude that 
Maxwell stresses in
the ICM were unimportant. That is probably a false impression, however.

While a meaningful rule of thumb,
the simple $\beta$ metric overlooks important details.
A particularly simple one is that the magnetic 
and gas pressure forces depend on $\ell_B$ and $\ell_P$, 
the spatial variation scales for B and P.
These scales need not be the same, especially in
a complex, mostly incompressible, turbulent setting. Then, when
$\beta >>1$, the magnetic field, being tangled and intermittent
in magnitude, may exert a force that is much stronger
than one would anticipate from a simple evaluation of $\beta$.
A better metric would be $\beta (\ell_B/\ell_P) \gg 1$.
This point has been made by \cite{oneil07} in a 3D MHD simulation study
of AGN jets in an ICM containing a tangled magnetic
field. Even for a mean $\beta \sim 100$, they found
that the ICM magnetic field had a very significant influence on
the subsonic expansion of the jet/ICM contact discontinuity, 
because local values of $\beta$ varied by factors of several on 
scales
that were much smaller than the relatively smooth ICM gas pressure.
Consequently magnetic pressure {\it gradients} were, in fact, sometimes
comparable to gas pressure {\it gradients}.

A second point is that the magnetic tension force, which resists bending of field
lines, is often more important than
the pressure force, especially in subsonic flows characteristic
of ICMs and many of their interfaces with AGN outflow residues. 
An example setting where this applies, mentioned in \S 2, is MHD turbulence, 
where magnetic tension distorts small scale eddies. The role of
magnetic tension can be estimated by comparing the
Maxwell stress in eq. \ref{navier} to the inertial, shear stress. 
From this we establish that magnetic tension may be important when $v_A \ge u_d (\ell_B/\ell_d)$.
If $\ell_B \sim \ell_d$, as is the case parallel to the field in
MHD turbulence, we recover the $v_A \ge u$ criterion from \S 2.
Other examples will be discussed below.

Another point to keep in mind, of course, is the possibility of
magnetic field amplification,
described by the induction eq. \ref{navier}.
Outside
reconnection regions where dissipation is clearly important,
the ideal MHD evolution of the field expressed in eq. \ref{navier}
consists of field compression and field line stretching. In complex
flows field line stretching is usually a bigger effect, since it
is limited only by the eventual backreaction of magnetic tension and 
local reconnection
when field lines ``cross''. For incompressible flow the ideal
version of eq. \ref{navier} reduces to $|B| = |B_0| \ell/\ell_0$, 
where $\ell$ measures the length of a field line segment. Thus, 
the field intensity is proportional to the stretching of the line.
This will vary stochastically in a turbulent flow, leading to
spatially intermittent field intensities.
The constraint imposed by magnetic tension, $v_A < u$, on the other hand,
assures that amplification of the magnetic
field will be limited to magnetic pressures, $P_B = (1/2) v_A^2 \rho$, less than the
turbulent, kinetic pressure, $P_k \sim u^2 \rho$. 

One very important ICM MHD concern
is the evolution of hydrodynamical
instabilities, particularly the Kelvin-Helmholtz (K-H) and
Rayleigh-Taylor (R-T) instabilities. These are both essentially
incompressible instabilities that occur along boundaries such as
the contact discontinuities separating AGN jet cocoons from the ICM
or relic radio bubbles from the ICM. 
Related, R-T-like instabilities can develop anywhere equilibrium pressure and 
density gradients take opposite signs, or equivalently where
pressure and entropy gradients have the same sign.
It turns out that surprisingly weak magnetic fields can significantly
modify these instabilities, sometimes quenching them and sometimes
controlling their nonlinear evolution even when magnetic stresses do not
prevent the instabilities from developing.

The K-H instability develops hydrodynamically
along a shear layer for wavelengths exceeding the shear layer
thickness, provided the shear velocity is not highly
supersonic.
Tension from a magnetic field component parallel to the velocity in the shear layer
and aligned with the wavevector of a perturbation
will resist growth of the instability, since growing oscillations
lengthen the boundary.
For a velocity discontinuity, $\Delta u$, separating fluids of
equal density the linear growth rate of a small perturbation with
wavenumber k is \cite{chand61}
\begin{equation}
\Gamma = (1/2) k \Delta u  \sqrt{(1 - 4 v_{A,||}^2/\Delta u^2)}.
\end{equation}
The perturbation
is stable ($\Gamma^2 < 0$) for $v_{A,||} \ge \Delta u/2$.
Linear stabilization of the K-H instability in the ICM
would roughly require $B_{\mu G} > 2 u_{d,100} n_{-2}^{1/2}$,
a condition that is likely to be met along interfaces of low density
and modest velocity contrast, as pointed out, for example,
by \cite{deyoung03}.

It is important to realize, however, even when it is initially much too weak to 
inhibit the linear K-H instability, that a magnetic
field can play a very important role in the nonlinear K-H instability.
This influence results from stretching (amplification) of the initial field within
the unstable flow.
For example, \cite{ryu00} carried out 3D MHD simulations of
K-H unstable flows, and found that even for initial
conditions with $v_{A,||} \sim 0.02 \Delta u$,
the unstable flow evolved into a relaxed, laminar form with aligned, self-organized
magnetic and velocity fields, contrary to the turbulent
velocity field that developed in the absence of a magnetic field.
The key insight is that the 3D hydrodynamical flow initially
forms into line vortices along the unstable slip surface, stretching
the magnetic field across each vortex. Subsequently, the
vortex is unstable in 3D to so-called rib
vortices transverse to the main vortex. This greatly adds to the
net stretching of the field, making it possible locally and temporarily
that $v_A > \Delta u$ and, thus for an initially weak field to
alter evolution of the flow. 

The linear R-T instability is influenced by a magnetic field 
parallel to a density discontinuity in a
way similar to the K-H instability; magnetic
tension resists the stretching of the boundary. Defining the
density jump across the discontinuity as $r = \rho_2/\rho_1$,
where region `2' is on top with respect to gravity, g, and 
letting $\tilde v_{A}^2 = B^2/[4\pi(\rho_1 + \rho_2)]$,
the growth rate is \cite{chand61} 
\begin{equation}
\Gamma = \sqrt{ [1/(r+1)]}\sqrt{[g k (r-1) - 2 \tilde v_{A,||}^2 k^2]}.
\end{equation}
In this case the influence of the magnetic tension
depends on wavelength, so that the instability is suppressed
when $k > k_0 = (1/2) g (r-1)/[\tilde v_{A,||}^2 (r + 1)]$, while
there is a maximum linear growth rate at wavenumber, $k_{max} = (1/2) k_0$,
given by 
\begin{equation}
\Gamma_{max} = (1/2\sqrt{2}) g (r-1)/[\tilde v_{A,||} (r+1)].
\end{equation}
As a mnemonic, imagine for a large density contrast, $r \gg 1$, that the
free fall over a time $\sim 1/\Gamma_{max}$ reaches a velocity $\sim \tilde v_{A,||}$, 
and a length $\sim 1/k_{max}$. 
It turns out even a vertical magnetic field
reduces linear growth of the R-T instability, although it does not
totally quench it. The asymptotic short wavelength growth rate is
of order $g/v_A$, compared to the hydrodynamical rate $\sim \sqrt{g k}$. 
These two rates are similar for $k \sim g/v_A^2$, mirroring the
horizontal field behavior.
As a rule of thumb, then, we can
expect magnetic inhibition of the R-T instability whenever the Alfv\'en
speed is greater than or comparable to $\sqrt{g \ell}$. That would
correspond roughly to a magnetic field $B_{\mu G} \ge 7 ((n_{-2}~g_{-7} \ell_{d.kpc})^{1/2}$, where $g_{-7}$ is the gravitational acceleration in
units of $10^{-7}~{\rm cm/s^{-2}}$.
Alternatively one could view this relation to indicate, given ICM magnetic fields
of order a microGauss, that R-T instabilities on scales smaller than
a few kpc are likely to be inhibited. As I noted earlier and others
have also emphasized \cite[e.g.,][]{rey05}, viscous effects on
this scale could also be important, depending on how much plasma
and magnetic fields reduce the effective free-streaming length for
protons.  Which influence, magnetic tension or viscosity, is dynamically
more important on this scale remains to be established \cite[e.g.,][]{kais05}.

Ideal MHD influences in the nonlinear R-T instability are a bit
complicated, as pointed out by \cite{jun95} in a 2D and 3D simulation study
of the nonlinear MHD R-T instability. The nonlinear hydrodynamic R-T
instability is characterized by the formation of dense fingers that
`drip' downward and light ``bubbles'' that rise. These eventually
lead to turbulent conditions. Jun et al. noted that
the vertical component of B aligned with gravity eventually dominates
the evolution of these structures, especially when the
initial field is only moderately weaker than needed to suppress the
linear instability. Then, in fact, since the vertical magnetic field
aligns with the edges of R-T fingers, it inhibits development
of a secondary R-T instability that otherwise disrupts those fingers.
Thus, it can actually enhance their early development. Ultimately, however,
even a relatively weak magnetic field caught up in the nonlinear
development of the R-T instability can be amplified through
stretching to restrict growth of turbulent behaviors.

There are a limited number of MHD simulation studies of AGN and/or
radio relic, bubble interactions with ICMs. They generally support
the points made here, including the realization that fields
at least several times weaker than those expected by the usual
metrics can play essential dynamical roles. As examples, \cite{rob04} and \cite{jones05}
considered 2D MHD buoyant bubbles in model ICMs with large scale
magnetic fields, pointing out that even when $\beta \gg1$ field stretching
could stabilize boundaries that otherwise were disrupted by R-T and
K-H instabilities. \cite{deyoung07} obtained similar results for
3D bubbles, but also noted that since field line tension acts only in the
plane containing the field and its curvature, the bubbles remained
subject to disruption along lines orthogonal to the field-gravity
plane. \cite{rus07} carried out 3D MHD bubble simulations 
in which the ICM magnetic field was tangled. They noted that
when the outer tangling scale was smaller than the size of the 
bubble the magnetic field was no longer effective in
stabilizing the bubbles. This result, confirmed by \cite{deyoung07}, reflects the fact that
disruption comes from instabilities on the scale of the
bubble. It is obvious that magnetic tension, the primary
MHD stabilizing influence, does not operate on scales
greater than the coherence length of the field. On those
scales we should expect the dynamics to be largely hydrodynamic, except
for magnetic pressure gradient effects. I already mentioned that
 \cite{oneil07}
pointed out how variations in magnetic pressure on scales smaller
than gas pressure variations can produce important dynamical
consequences even when $\beta \gg 1$.

\section{Conclusion}

As a number of previous authors have discussed, Coulomb mean free
paths in the ICM can sometimes be uncomfortably large when one wants to model
the ICM as an ideal fluid. Yet observed features, such as shocks,
sharp contact discontinuities and turbulence strongly suggest continuum,
fluid behaviors. It seems likely that a combination of
plasma instabilities and magnetic field influences reduce the
effective particle free-streaming lengths enough to allow
relatively inviscid, fluid-like behaviors at least beyond kiloparsec 
scales.  Whether
these effects can also reduce diffusion rates of 
electrons sufficiently to effectively quench thermal conduction
is less certain. In any case electrical conductivity is very likely
to be sufficient to apply a frozen-in magnetic field model on
these scales. 

Even though magnetic pressures in the ICM are likely to be
much smaller than gas pressures, magnetic fields
there can still play a significant dynamical role. This is especially because
the local Alfv\'en velocity can be comparable to or exceed
local flow speeds, since ICM flows are usually subsonic. In that
case magnetic tension forces can compete with dynamical
stresses. This probably influences ICM turbulence on small, say 
kiloparsec, scales and helps stabilize flows that are otherwise 
hydrodynamically unstable.

\acknowledgements 
This work was supported in part by NSF grant AST 06-07674 to the University of
Minnesota and by the Minnesota Supercomputing Institute. I am grateful
to a number of collaborators over the years who contributed
insights and results that have gone into this review, especially
Dave De Young, Adam Frank, Francesco Miniati, Sean O'Neill and Dongsu Ryu.

%%% THE BIBLIOGRAPHY
%%%
%%% CONSULT SECTION 3 OF "INSTRUCTIONS FOR AUTHORS" FOR HOW TO USE NATBIB.
%%% AUTHORS ARE ENCOURAGED TO USE EITHER THE "THEBIBLIOGRAPY" ENVIRONMENT
%%% BY UNCOMMENTING (DELETING THE "%" SYMBOL) THE COMMANDS BELOW, OR BY
%%% USING THE BIBTEX ENVIRONMENT. TO FIND OUT WHICH IS APPLICABLE TO YOUR
%%% CONTRIBUTION, CONSULT THE VOLUME EDITORS FOR YOUR PROCEEDINGS.
%%%


\begin{thebibliography}{}
\bibitem[Beck et al. 2006]{beck}
Beck, R., Brunetti, G. \& Feretti, L. 2006, Astron. Nachr., 327, 385
\bibitem[Bertschinger \& Meiksin 1986]{bert86}
Bertschinger, E. \& Meiksin, A. 1986, Ap. J., 306, L1
%\bibitem[Biskamp 1993]{bisk93}
%Biskamp, D. 1993, Nonlinear Magnetohydrodynamics (Cambridge: Cambridge University Press)
\bibitem[Boyd \& Sanderson 2003]{boyd03} 
Boyd, T. J. M. \& Sanderson, J. J. 2003, The Physics of Plasmas (Cambridge: Cambridge University Press)
\bibitem[Braginksii 1965]{brag65}
Braginskii, S. I. 1965, Rev. Plasma Phys., 1, 205
\bibitem[Brunetti \& Lazarian 2007]{brun07}
Brunetti, G. \& Lazarian, A. 2007, M. N. R. A. S., 378, 245
\bibitem[Carilli \& Taylor 2002]{car02}
Carilli, C. L. \& Taylor, G. B. 2002, A.R.A.A., 40, 319
\bibitem[Chandrasehkar 1961]{chand61}
Chandrasekhar, S. 1961, Hydrodynamic and Hydromagnetic Stability (London: Oxford University Press)
\bibitem[Croston et al. 2005]{croston05}
Croston J. H., Hardcastle, M. J., Harris, D. E., Belsole, E., Birkinshaw, M. \& Worrall, D. M. 2005, Ap.J., 626, 733
\bibitem[De Young 2003]{deyoung03}
De Young, D. S. 2003, M. R. A. S., 343, 719
\bibitem[De Young et al. 2007]{deyoung07}
De Young, D. S., O'Neill, S. M. \& Jones, T. W. 2007 (in preparation)
\bibitem[Dolag 2006]{dolag06}
Dolag, K. 2006, Astron. Nachr., 327, 575
\bibitem[Fabian et al. 2005]{fab05}
Fabian, A. C., Reynolds, C. S., Taylor, G. B. \& Dunn, R. J. H. 2005, M. N. R. A. S., 363, 891
\bibitem[Goedbloed \& Poedts 2004]{goed04}
Goedbloed, H. \& Poedts, S. 2004, Principles of Magnetohydrodynamics (Cambridge: Cambridge University Press)
\bibitem[Gruzinov 2002]{gruz02}
Gruzinov, A. 2002, arXiv:astro-ph/0203031
\bibitem[Gruzinov 2006]{gruz06}
Gruzinov, A. 2006, arXiv:astro-ph/0611243
\bibitem[Hall 1980]{hall80}
Hall, A. N. 1980, M. N. R. A. S., 190, 353
\bibitem[Jones \& De Young 2005]{jones05}
Jones, T. W. \& De Young, D. S. 2005, Ap. J., 624, 586
\bibitem[Jun et al. 1995]{jun95}
Jun, B.-I., Norman, M. L. \& Stone, J. M. 1995, Ap. J., 453, 332
\bibitem[Kaiser et al. 2005]{kais05}
Kaiser, C. R., Pavlovski, G., Pope, E. C. D. \& Fangohr, H. 2005, M. N. R. A. S., 359, 493
\bibitem[Kronberg 1996]{kron96}
Kronberg, P. P. 1996, Space Sci. Rev., 75, 387
\bibitem[Kulsrud 2004]{kuls04}
Kulsrud, R. M. 2004, Plasma Physics for Astrophysics (Princeton: Princeton University Press)
\bibitem[Lazarian 2007]{laz07}
Lazarian, A. 2007, arXiv:0707.0702
\bibitem[Narayan \& Medvedev 2001]{nar01}
Narayan, R. \& Medvedev, M. V. 2001, Ap.J. Lett., 562, L129
\bibitem[O'Neill et al. 2007]{oneil07}
O'Neill, S. M., Jones, T. W. \& Ryu, D. 2007 (in preparation)
\bibitem[Pope et al. 2005]{pope05}
Pope, E. C. D.,Pavlovski, G., Kaiser, C. R. \& Fangohr, H. 2005, M. N. R. A. S., 364, 13
\bibitem[Priest \& Forbes 2000]{priest00}
Priest, E. \&  Forbes, T. 2000, Magnetic Reconnection: MHD Theory and Its
Applications (Cambridge: Cambridge University Press)
\bibitem[Reynolds et al. 2005]{rey05}
Reynolds, C. S., McKernan, B., Fabian, A. C., Stone, J. M. \& Vernaleo J. C. 2005, M. N. R. A. S., 357, 242
\bibitem[Robinson et al. 2004]{rob04}
Robinson, K., et al. 2004, Ap. J., 601, 621
\bibitem[Ruszkowski et al. 2007]{rus07}
Ruszkowski, M., Ensslin, T. A., Brueggen, M., Heinz, S. \& Pfrommer, C.
2007, M. N. R. A. S., 378, 662
\bibitem[Ryu et al. 2000]{ryu00}
Ryu, D., Jones, T. W. \& Frank, A. 2000, A. J., 545, 475
\bibitem[Ryu et al. 2007]{ryu07}
Ryu, D., Kang, H., Cho, J. \& Das S. 2007, (in preparation)
\bibitem[Schekochihin et al. 2005]{schek05}
Schekochihin, A. A., Cowley, S. C., Kulsrud, R. M., Hammett, G. W. \& Sharma, P. 2005, Ap.J., 629, 139
\bibitem[Schuecker et al. 2004]{schueck04}
Schuecker, P., Finoguenov, A., Miniati, F., Boehringer, H. \& Briel, U. G. 2004, A\&A, 426, 387
\bibitem[Spitzer 1962]{spitz62}
Spitzer, L. 1962, Physics of Fully Ionized Gases (New York: Interscience Publishers)
\bibitem[Stix 1992]{stix92}
Stix, T. H. 1992, Waves in Plasmas (New York: AIP)
\bibitem[Vikhlinin et al. 2001]{vik01}
Vikhlinin, A., Markevitch, M. \& Murray, S. S. 2001, Ap.J. Lett., 549, L47
%\bibitem[Vikhlinin et al. 2001b]{vik01b}
%Vikhlinin, A., Markevitch, M., Forman, W. \& Jones, C. 2001, Ap.J. Lett., 555, L87
\bibitem[Worrall \& Birkinshaw 2000]{worr00}
Worrall, D. M. \& Birkinshaw, M. 2000, Ap.J., 530, 719
%\bibitem[]{}
%\bibitem[]{}
%\bibitem[]{}
%\bibitem[]{}
%\bibitem[]{}
\end{thebibliography}
\end{document}